\documentclass[11pt,a4paper]{article}
\usepackage{jheppub,amsmath,  amssymb,slashed,url,bm,textgreek,upgreek}
\usepackage{graphicx}
\usepackage{epstopdf}
 
\def\tt{{\mathfrak t}}

\def\a{{\sf a}}

\def\be{\begin{equation}}
\def\ee{\end{equation}}

\def\hat{\widehat}

\def\O{{\mathcal O}}

\def\A{{\mathcal A}}
\def\d{{\mathrm d}}

\def\b{{\sf b}}
\def\R{{\mathbb R}}

\def\[{\bigl [}

\def\]{\bigr ]}

\def\H{{\mathcal H}}

\font\teneurm=eurm10 \font\seveneurm=eurm7  \font\fiveeurm=eurm5
\newfam\eurmfam
\textfont\eurmfam=\teneurm \scriptfont\eurmfam=\seveneurm
\scriptscriptfont\eurmfam=\fiveeurm

\font\teneusm=eusm10 \font\seveneusm=eusm7 \font\fiveeusm=eusm5
\newfam\eusmfam
\textfont\eusmfam=\teneusm \scriptfont\eusmfam=\seveneusm
\scriptscriptfont\eusmfam=\fiveeusm

\font\tencmmib=cmmib10 \skewchar\tencmmib='177
\font\sevencmmib=cmmib7 \skewchar\sevencmmib='177
\font\fivecmmib=cmmib5 \skewchar\fivecmmib='177
\newfam\cmmibfam
\textfont\cmmibfam=\tencmmib \scriptfont\cmmibfam=\sevencmmib
\scriptscriptfont\cmmibfam=\fivecmmib

\title{An Observer in de Sitter Space, and Rereading Everett}

 \author{Edward Witten}
\affiliation{School of Natural Sciences, Institute for Advanced Study,\\ 1 Einstein Drive, Princeton, NJ 08540 USA}
\abstract{After re-examining a classic article by Hugh Everett, I will discuss the entropy of de Sitter space from the standpoint of an observer living in the
space.  This provides a nice example of the importance -- for certain questions  -- of taking seriously the fact that  in cosmology any observer is really part of the
universe that is being observed.  (This article has appeared in a special issue of International Journal of Modern Physics A in honor of Gerard 't Hooft.)
}
\def\vis{{\rm{vis}}}
\def\tot{{\rm{tot}}}
\def\dS{{\rm{dS}}}
\def\Tr{{\rm{Tr}}}
\def\obs{{\rm{obs}}}
\def\max{{\rm{max}}}
\def\NB{{\rm{NB}}}
\def\i{{\rm i}}
\begin{document}\maketitle

\section{Rereading Everett}\label{everett}

In nonrelativistic quantum mechanics, we usually consider the observer to be probing a quantum system from outside,
but of course in the real world, any observer is part of a larger quantum system.      That is most conspicuous in cosmology -- especially but
not only in the case of a closed universe.    My main goal here, following some relatively recent  papers \cite{CLPW,W},
 is to explain that taking seriously the observer as part of the quantum system
can  lead to useful results, not just platitudes.

\begin{figure}\label{one}
 \begin{center}
   \includegraphics[width=5.9in]{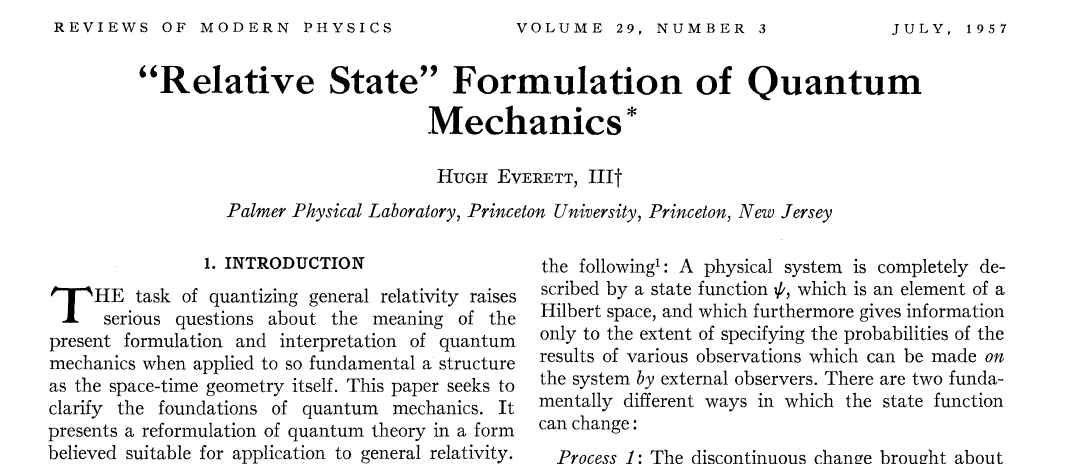}
 \end{center}\caption{{\small A classic article by Hugh  Everett.}}
\end{figure}

But first I want to re-examine a famous paper by Hugh Everett from 69 years ago that was motivated by the idea of the observer as part of the
quantum system (see fig. \ref{one}).   Everett motivated his analysis with the claim  that general relativity requires a rethinking of quantum mechanics.
``The task of quantizing general relativity,'' he begins, ``raises serious questions about the meaning of the present formulation of quantum mechanics when applied to
so fundamental a structure as the spacetime geometry itself.   This paper $\dots $ [presents] a reformulation of quantum theory in a form believed suitable
for application to general relativity.''

Elaborating on that first paragraph, Everett writes, ``How is one to apply the conventional
formulation of quantum mechanics to the space-time geometry itself?   The issue becomes 
especially acute in the case of a closed universe.  There is no place to stand outside the system
to observe it.   There is nothing outside it to produce transitions from one state to
another.... No way is evident to apply the conventional formulation of quantum mechanics to
a system that is not subject to {\it external} observation.''   
By external observation, Everett means observation by an observer outside the system. Everett wanted to consider an observer who is part of the quantum system.

A quick explanation of the difference between  Everett and a classic interpretation of quantum mechanics as advocated by Niels Bohr is the following. 
Bohr said that when an observation is made, the result is recorded in the measuring device.     Then
the observer is supposed to know the result recorded by the measuring device.    Bohr did not specify
the sense in which the observer knows that result.   But it is not supposed to be by measuring the measuring device
to learn its state, since that would lead to an obvious infinite regress.   Instead the observer was to know the state of the measuring device in
some undefined sense.

Everett instead said that when an observation is made and the observer looks at the measuring device to learn the result,
a quantum mechanical entanglement is created between the outcome of the measurement and the state of the observer's memory. (Everett does not use the term
``entanglement'' but describes the idea.   The ``relative state'' of the title of the article is the density matrix of the observer's memory.)  
Then, the observer is supposed to know the state of the observer's memory.    Everett did not specify the sense in which the observer
knows the state of the observer's memory.     But this is certainly  not supposed to be by measuring the state of the memory, since that would lead
back to the same infinite regress that Everett was trying to avoid.  

When I first read the paper, possibly 15 years ago, I was completely dissatisfied and could not believe that anyone thought that Everett had made a fundamental
contribution to understanding quantum mechanics.   It seemed that Everett had just shifted the problem from the measuring device to the observer's memory, 
more exactly ``in what sense does the observer know the state of the observer's memory?'' as 
opposed to ``in what sense does the observer know the state of the measuring device?''

Since this is a fairly obvious criticism, did Everett try to address it?   First, Everett did not assume an observer is made of flesh and blood; he also contemplated
computers or other artificial observers.    ``As models for observers we can, if we wish, consider automatically functioning machines, possessing sensory apparatus and
coupled to recording devices ... constructed so that its present actions shall be determined not only by its present sensory data, but by the contents of its 
memory as well.''    For such a machine, Everett gives a criterion for {\it another} observer to decide that it knows something:
``For such machines we are justified in using such phrases as `the machine has perceived $A$' or `the machine is aware of $A$' if the occurrence of $A$ is
represented in the memory, since the future behavior of the machine will be based upon the occurrence of $A$.''

{In other words, Everett gives a criterion for {\it another} observer to decide that the first observer knows something.     To me, this is yet another version of the
familiar infinite regress.}

{As I have said, when I first read the article, I was quite dissatisfied.      But over the years, I have had a few second thoughts.   One thought is that I do not
know what is happening in our brains when we think we know something.      Is it conceivable that Everett is actually correct as a matter of biology and physics? 
Do we directly ``know'' the state of our memory as opposed to ``measuring'' it?   Either a ``yes'' or ``no'' answer to that question seems problematical.  (``No'' seems to lead back to the
infinite regress and ``yes'' seems to claim that the brain somehow transcends the laws of physics.)  
I am afraid I do not have anything useful to say.}

Because both ``yes'' and ``no'' seem unsatisfying, I have also considered 
the following.   In mathematics, there are postulates or assumptions but there are also undefined terms.     For example, in Euclidean
geometry, one does not define the concept of a ``point'' or a ``line'' but one postulates properties that they satisfy.       Maybe in the foundations of quantum mechanics,
we should take ``knowing'' as an undefined term.   It does seem that the meaning of ``knowing'' is the main difficulty.

Anyway my view is that Everett did not ``solve'' a ``measurement problem'' but reading his article helps in seeing the essential problem --  the meaning of ``knowing.'' 

While Everett centered the interpretation of quantum mechanics on the perspective of an observer, this has been done even more sharply by some
later authors.   
I recommend the article \cite{FMS} by Fuchs, Mermin, and Schack.

\section{An Observer in de Sitter Space}

But all this can sound like platitudes.  
 I want to move on to an area where we can learn something concrete and useful by taking seriously that the observer is part of the system.   This will involve
 quantum mechanics in a de Sitter-like spacetime.   But first I will explain some basics about de Sitter space.
 
 De Sitter
space is the maximally symmetric solution of Einstein's equations with a positive cosmological constant:
 \be\label{first}\d s^2 =-\d t^2+ R^2\cosh^2 (t/R) \d\Omega^2, \ee
 where $\d\Omega^2$ is the metric of a unit three-sphere.    Thus, this is a closed universe.   The radius of the sphere at time $ t$ is $R\cosh(t/R)$ and so  the
 sphere expands exponentially towards both the past and the future.    The future-going part of this expansion is the input to the theory of cosmic inflation, and may also
 be a good description of what is beginning to happen in the current universe.
 
 Classically, because of the exponential expansion, the universe becomes almost everywhere perfectly
   empty in the far future (the quantum story is more subtle as we will discuss).
   That the expanding universe becomes empty 
 is actually a key input in the theory of cosmic inflation:   since the universe quickly becomes everywhere virtually empty, the memory of the initial state is lost
 and it is possible to make a definite prediction about the future.

  \begin{figure}
 \begin{center}\label{two}
   \includegraphics[width=3in]{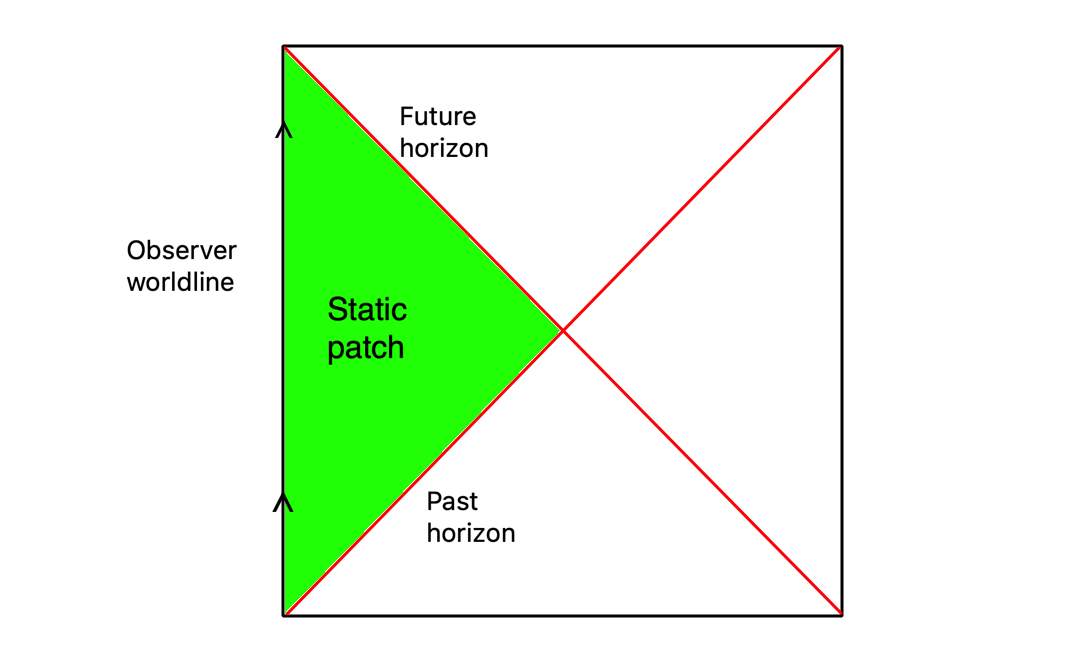}
 \end{center}\caption{{\footnotesize A Penrose diagram of de Sitter space.   Time runs vertically; the bottom and top of the diagram correspond respectively to $t=-\infty$
 and $t=+\infty$, where $t$ is defined in eqn.  (\ref{first}). Plotted from left to right is, roughly, the latitude on the sphere.   The left edge of the diagram is
 a geodesic describing an observer who remains at all time at the south pole of the sphere.   In the diagram, causal signals propagate at an angle no more than
 $\pi/4$ from the vertical, so  the spacetime regions that the observer can see and influence are bounded by future and past horizons, as shown.  Shaded in green is the region that the observer can both see and influence.   It is called the static patch, because in a suitable coordinate system, it has a time-independent metric. } }
\end{figure}

 Up to the symmetries of de Sitter space, any geodesic is equivalent to a geodesic that remains for all time at the south pole of the sphere. We will regard
 this geodesic as the world line of an observer, at least in a semiclassical approximation.     In drawing a Penrose
 diagram for de Sitter space, coordinates can be chosen so that this chosen geodesic is the left edge of the diagram, as shown in fig. \ref{two}.   Since causal signals can propagate at most
 at a $\pi/4$ angle from the vertical in the Penrose diagram, the regions of spacetime that the observer can see or influence are bounded by future and past
 horizons, as was first recognized by Willem de Sitter around 1917.   The region that the observer can both see and influence is sketched in green in the figure
 and is known as a ``static patch.''
      The reason for the name  is that coordinates can
be chosen in which the metric of this region  is time-independent.

Bekenstein and Hawking famously discovered in 1972-4 that the area $A$ of a black hole horizon, or more precisely the combination $A/4G$ where $G $ is Newton's constant,
can be interpreted as a sort of entropy. 
Gibbons and Hawking \cite{GH} proposed a few years later 
 that also in the case of a cosmological horizon, $A/4G$ should be interpreted as an entropy, which they said ``measures the lack of information of the
observer about the regions that he cannot see.''

The horizon area of de Sitter space is actually $A=4\pi R^2$ (where $R$ is the parameter in the metric (\ref{first})),  regardless of the time at which 
one measures it.  This actually follows from the symmetries of de Sitter space.
In the real world -- assuming that we are now entering a phase of de Sitter like expansion -- the de Sitter entropy is very big, roughly $10^{120}$.

The total entropy of the observable universe would then be the ``visible'' entropy $S_{\vis}$ of matter, radiation, and black holes in the visible part of the universe plus the entropy of the de Sitter horizon: \be S=\frac{A}{4G}+S_\vis.\ee
Some numbers:  in the real world, the entropy of all the stars in the observable universe is roughly $10^{84}$, the entropy of the cosmic microwave radiation is about $10^{89}$, the entropy of
super-massive black holes in the centers of galaxies is about $10^{104}$, and as already noted, the entropy of the cosmological horizon is about $10^{120}$.

{The meaning of the de Sitter entropy has been even more obscure than the black hole entropy of Bekenstein and Hawking.    Entropy is supposed to, in some sense, 
count the number of quantum states.     In the case of a black hole, although we do not have a full understanding, there are  at least  some situations in which we can
explain an entropy $A/4G$ by counting states:    \begin{itemize}\item  
In the case of a supersymmetric black hole,
we can directly count the quantum states (following classic work of Strominger and Vafa \cite{SV} in 1996,  and a whole series of more general and increasingly precise results since then).  \item  In the AdS/CFT correspondence, at high temperatures, almost every state is a black hole state (with a mass
determined by the temperature)
and the black hole entropy is just the thermal entropy of the CFT.\end{itemize}
By contrast, for de Sitter space (or more general cosmological horizons) there is no satisfying analog of any of that, so the de Sitter entropy has been more obscure.}

Even though it has been difficult to give a microscopic interpretation of the de Sitter entropy, an encouraging  fact has been that the total entropy understood as
$$S_\tot=\frac{A}{4G} +S_{\vis} $$
does appear to obey a Second Law, that is, \be \frac{\d S_\tot}{\d t}\geq 0.\ee
In the classical limit, this statement is the Hawking area theorem \cite{Hawking}; a version that includes the first
quantum correction was proved by  Wall \cite{Wall}.

 \begin{figure}
 \begin{center}\label{three}
   \includegraphics[width=2.2in]{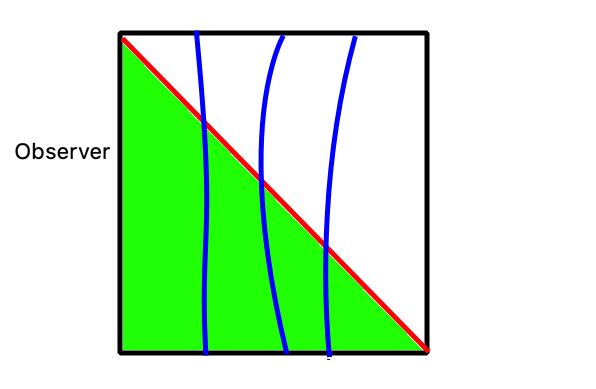}
 \end{center}\caption{{\footnotesize Shaded in green is the region of spacetime that is visible to an observer whose worldline is the left edge of the Penrose
 diagram.   This region, which is bounded by the future cosmological horizon, 
 constitutes a sort of open system, since particles and other objects can freely leave it,
 as shown. }}
\end{figure}

Of course (fig. 3), particles (and stars, and black holes) can cross the cosmological horizon and disappear from view.   
When particles cross the cosmological horizon, $S_\vis$ goes down, but according to the Second Law,  $A/4G$ increases by more, and $S_\tot$ increases.

Even without a microscopic definition of entropy, a fundamental observation was made by Bousso \cite{Bousso}  on purely thermodynamic grounds.      As I remarked before, de Sitter
space becomes empty in the far future (and this is fundamental in the theory of cosmic inflation).   Since de Sitter space, from any starting point, 
spontaneously evolves to be empty, the Second Law
seems to imply that ``empty de Sitter space'' must be a state of maximum entropy.   Again, this was proposed purely based on thermodynamics, with no microscopic definition of
entropy.     It can be partly confirmed by showing that placing a black hole inside the cosmological horizon actually reduces $S_\tot$
\cite{MKNI}.
  Adding a black hole ``pulls in'' the cosmological horizon, making $S_\tot$ smaller.

Actually, at the quantum level, ``empty de Sitter space'' is a subtle concept.   By ``empty Minkowski space,'' we would mean the vacuum state, the state that
has lowest energy in every Lorentz frame.   Such a state can exist because in Minkowski space, the energy is the conserved charge associated
with a Killing vector field that is everywhere future-directed timelike.    De Sitter space has a large symmetry group $SO(4,1)$, just as large as the Poincar\'{e} symmetry of Minkowski space.  But none
of the corresponding Killing vector fields is everywhere future-directed timelike, so in the absence of gravity there is no positive-definite conserved charge for physics in de Sitter space.   There is nothing we can minimize to construct a de Sitter ``vacuum state.''

In de Sitter space, in a sense  the  closest one can come to finding an everywhere future-directed timelike Killing
vector field is the following.
For any geodesic observer, such as the one on the left edge of the Penrose diagram of fig. 2, there is  a Killing vector field that is future-directed timelike in the region causally accessible to the observer -- the static patch -- 
but not in other regions.
  The corresponding conserved charge, which I will call $H$, will play an important role.   The observer  might regard this
as the ``Hamiltonian.''     However, while $H$ is, at the classical level, positive for excitations in the static patch, it is negative for excitations in the region that
is spacelike separated from the static patch, for in that region the Killing vector field associated to $H$ is past-directed timelike.

With gravity included, matters are worse.  Remember that in an asymptotically Minkowski spacetime, in the presence of gravity, the energy, momentum, and angular momentum
are all defined as surface integrals at spatial infinity.    Since de Sitter space is a closed universe, there is no spatial infinity and no way to define conserved charges associated to
spacetime symmetries.     Rather, after including gravity, the symmetry generators of de Sitter space become constraint operators; they commute with physical operators and annihilate physical states.  
So there really is no conserved quantity such as energy that a state could minimize.

But a quantum field in a background de Sitter space  does have a distinguished de Sitter invariant state $S_\dS$
(sometimes called the Bunch-Davies state).    It can be defined by analytic continuation from a four-sphere.     Starting with the de Sitter metric 
$\d s^2=-\d t^2+R^2 \cosh^2(t/R) \d\Omega^2$
and continuing to Euclidean signature by $t\to \i t_{\sf E}$, we arrive at a Euclidean signature metric 
\be \d s_E^2=\d t_{\sf E}^2+ R^2\cos^2(t_{\sf E}/R)\d\Omega^2 \ee
that actually describes a four-sphere of radius $R$.      Since the four-sphere has no boundary, correlation functions of a 
quantum field on the four-sphere can be uniquely defined
with no choice of boundary condition.     Continuing back to Lorentz signature, these correlation functions become correlation functions in the natural de Sitter invariant
state $\Psi_\dS$.  

The state $\Psi_\dS$ has the remarkable property that correlation functions in this state are, in a certain sense, thermal at the de Sitter temperature $T=1/2\pi R$
\cite{FHN,GH}.     To be precise, consider a geodesic observer in de Sitter space; we may as well consider
an observer whose worldline is the left edge of the Penrose diagram of fig. \ref{two}.    Such an observer is forever at rest at the south pole of the sphere, and the
time coordinate $t$ in eqn. (\ref{first}) measures the proper time elapsed along the observer's worldline.

  \begin{figure}
 \begin{center}\label{orbit}
   \includegraphics[width=2.2in]{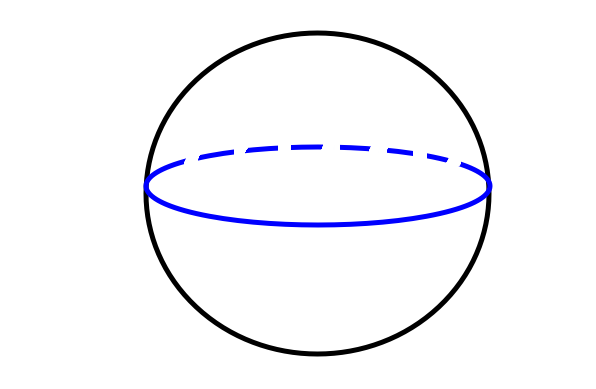}\caption{{\footnotesize A geodesic in Lorentz signature analytically continues in Euclidean signature to a great circle on the 
   sphere, such as the one shown in blue.}}
 \end{center}\end{figure}

 We assume that what the observer measures are correlation functions along the observer's worldline, say $\langle \phi(t)\phi(t')\rangle$ where $\phi$ is a quantum field (such as a 
 component of the electromagnetic field).   In Euclidean signature, the observer's worldline analytically continues to a great circle parametrized by $t_{\sf E}$.
The great circle circumference is $2\pi R$ so $t_{\sf E}$ is a periodic variable with period $2\pi R$. 

In other words, the correlation functions that the observer measures can be analytically continued to imaginary time, and after doing so, the imaginary
time coordinate is a periodic variable with  period $2\pi R$.
   In quantum mechanics, periodicity of the  imaginary time coordinate  is the hallmark of thermal correlation functions.      
   In the present case, the periodicity means that the observer sees thermal correlation functions at 
 a temperature $T_\dS=1/2\pi R$ that is inverse to the period.   $T_\dS$  is  known as the de Sitter temperature.   It is the de Sitter analog of the
Hawking temperature of a black hole.     Thus, the state $\Psi_\dS$, which is the closest de Sitter analog of the Minkowski space vacuum, 
and thus represents the quantum version of ``empty de Sitter space,''  actually has thermal properties at the de Sitter temperature.    
Part of what makes this possible is that the state 
$\Psi_\dS$ was not defined by minimizing anything and so is not guaranteed to describe a truly ``empty'' universe in any naive sense.

{Now that we understand the context, let us  try to understand what might be meant by the entropy of the region causally accessible to an observer (the green region in fig. \ref{two}).
    This region has been called a ``static patch''  because it is invariant under the action of one of the de Sitter symmetry generators, namely the generator $H$ that was introduced earlier.      $H$ moves the observer worldline forward in proper time, in the sense that along this worldline,  $H$ acts as $ \frac{\d}{\d\tau}$, where $\tau$ is the proper time of the observer.   $H$
 can be regarded roughly as the Hamiltonian of the static patch. 
 
 Usually in quantum theory, to define the entropy of a system, we start with the density matrix $\rho$ describing the system -- for example a thermal density matrix
 $\rho=\frac{1}{Z}e^{-\beta H}$.      Then, following von Neumann, the entropy is defined in terms of the density
 matrix as 
 \be S=-\Tr\,\rho\log\rho.\ee
 
 What is a  density matrix?    The density matrix of a system is usually an operator  that acts on the Hilbert space of the system.  That is not quite the right notion  here, because
there is no such thing as ``the Hilbert space of the static patch.''     One can define a Hilbert space $\mathcal H_\dS$ of the whole universe, but there is
 no Hilbert space that describes just the degrees of freedom in the static patch.    Technically, an attempt to define a Hilbert space of the static patch fails because of ultraviolet divergences near the horizon.
 
 Although the observer does not have a Hilbert space, the observer does presumably have access to observables -- like the electromagnetic field along the observer worldline -- that can be measured.
    These observables are operators on the full Hilbert space $\mathcal H_\dS$ of the theory, which describes degrees of freedom in the static patch and also degrees of freedom beyond the
 horizon that the observer cannot see or influence.  Observables can be added or multiplied, so  they form  an ``algebra.'' 
 
   {\it If}  the observer's algebra $\A$
 has a ``trace function,''  then it is possible to define density matrices and entropies from the observer's point of view, even though the observer does not have
 a Hilbert space.  A trace function is simply a complex-valued linear function
  $\a \to \Tr\,\a$ that is defined for any $\a\in\A$ and satisfies 
 \be\Tr\,\a\b=\Tr\,\b\a \ee
 for any $\a,\b\in\A$,
along with $\Tr\,\a^\dagger = \overline{\Tr\,\a}$, and a  condition of nondegeneracy.      We just define the density matrix $\rho$ of a state $\Psi\in\mathcal H$ by the usual relation
 \be\langle \Psi|\a|\Psi\rangle = \Tr\,\a\rho.\ee
 An operator $\rho\in \A$  (or more precisely, in general an operator $\rho$ affiliated to $\A$) that satisfies this condition exists and is unique because of nondegeneracy of the trace.   Once a density matrix is defined, entropy can be defined by the usual formula
 \be S=-\Tr\,\rho\log\rho.\ee

 {We want to implement this idea for the case that the algebra $\A$ is the algebra of all operators along the observer worldline.    In general the observer
 worldline is a timelike curve.   A nice fact about quantum field theory (in the absence of gravity or with gravity treated perturbatively) is that it is indeed possible to define
 the algebra of operators along a {\it timelike} worldline such as the worldline of an observer.
 Because of ultraviolet divergences -- analogous to the divergences that prevent us from defining a Hilbert space of the static patch -- it is generically {\it not} possible to define an algebra
 of quantum field operators along a {\it spacelike} path.    Happily the observer's worldline is always timelike so there is no difficulty.
 
 If we do precisely what I have said so far, we run into a snag.    In the absence of gravity, this algebra can be defined but does not have a trace function so we cannot define
 entropies.     In the presence of gravity, something happens that is worse: we cannot really define an algebra precisely in the way that I have said. 
 That is because in the presence of gravity, the symmetries of de Sitter space become constraints.
In more detail, as noted previously, the de Sitter generator $H$ is a symmetry of de Sitter space together with the observer.   In the presence of gravity, $H$ becomes a constraint operator and only operators $\mathcal O$ that commute with $H$ are physically meaningful.
   Since $H$ is derived from the vector field that along the worldline acts as  $\frac{\d}{\d\tau}$, a quantum field operator $\phi(\tau)$   -- or in more detail $\phi(x^\mu(\tau))$, where $x^\mu(\tau)$ is the 
 position of the observer
 at proper time $\tau$ --  does not satisfy the constraint. 
 Rather
 \be [H,\O]=-\i\frac{\d\O}{\d\tau}.\ee
 
 One way to get operators that commute with $H$ is to consider not $\phi(\tau)$ but its integral over the whole worldline (for $\phi$ such that this integral converges):
 \be \phi \to \int_{-\infty}^\infty \d \tau\, \phi(\tau). \ee
   But we probably do not want to restrict the observer to considering only such integrated operators.  Measuring such an operator would take a long time!
     A better idea is, roughly, to equip the observer with a clock and let the
 observer measure quantum fields as a function of the time told by the clock.      We give the observer a Hilbert space $\H_\obs$ of functions $L^2(\R_+)$ of a non-negative variable $q$,
 and a Hamiltonian
 \be H_\obs= mc^2+q \ee that acts on this Hilbert space.
 Here $m$ is the observer mass (we take $q\geq 0$ so that the observer system has minimum energy $mc^2$).     We define an extended Hilbert space
 $\H_\tot=\H_{\dS}\otimes H_\obs$ which describes the states of the de Sitter space together with the observer.
 
 We have to include $H_\obs$ as an additional contribution to the constraint operator, which now becomes
 \be H_{\tot}=H+H_\obs.\ee    We can now give an improved definition of what we mean by the observer algebra $\A$:   it is the algebra of operators along the observer worldline
 (acting on $\H_\tot=\H_\dS\otimes \H_\obs$) that commute with the extended constraint $H_\tot$.      Though the details will not be important in this article, it
  turns out that one can describe $\A$ explicitly: it is generated by $q$ itself along with operators
 $$\hat\phi_s =\Pi \phi(p+s)\Pi,$$
 where $\phi$ is any quantum field of the theory, $p=-\i \frac{\d}{\d q}$ is conjugate to $q$, $\Pi$ projects on $q\geq 0$, and $s\in \R$.
 
 Remarkably, the algebra defined this way does have a trace function and therefore it is possible to define density matrices and entropies.   To motivate the construction, let me recall something about
 ordinary quantum mechanics.   For a quantum system with an $N$-dimensional Hilbert space, a mixed state of maximum entropy is   a ``maximally mixed''  state
 whose density matrix is a multiple of the identity operator $\rm{Id}$: \be\rho_\max =  \frac{\mathrm{Id}}{N}.\ee
   Hence up to a multiplicative constant, the trace of an operator $\a$ is the same as its expectation $\Tr\,\a\rho_\max$ in a maximum entropy state:
 \be\Tr\,\a=\Tr\,\a\,{\rm{Id}}=N \,\Tr\,\a\rho_\max.\ee
 
 {This suggests that for the algebra $\A$ that describes the observations of a de Sitter observer, we should define a trace in terms of the expectation in a maximal entropy state.
   But what is a state of maximum entropy?  Repeating Bousso's argument \cite{Bousso}, but now in the presence of an observer, 
   we would say that
 \begin{itemize}\item In the far future, the static patch becomes ``empty'' except for the presence of the observer;
    \item In the far future, the observer comes into equilibrium with the quantum fields, meaning that the observer energy is thermally distributed at the de Sitter
    temperature. \end{itemize}}
    
    {Assuming then that 
   the state just described to which the static patch and the observer spontaneously evolve must have maximum entropy,  
 we conclude  that a state of maximum entropy must be
    \be\Psi_\max=\Psi_\dS \cdot e^{-\beta_\dS q/2}\sqrt{\beta_\dS},~~~~~~~~\beta_\dS=2\pi R=\frac{1}{T_\dS}. \ee
     It is actually true that the expectation in the state $\Psi_\max$ behaves as a trace:
    $$\langle \Psi_\max|\a\,\b|\Psi_\max\rangle=\langle \Psi_\max|\b\,\a|\Psi_\max\rangle.$$
       That is hopefully a surprising result, but it can be proved by a short, elementary computation starting with the thermal property of $\Psi_\dS$; see \cite{W}.
  This result actually essentially goes back to the early 1970's.  It was discovered by mathematicians M. Takesaki and A. Connes
in their work on  von Neumann algebras.
      
    {Having defined a trace, we can also define density matrices and entropies.   The resulting definition  has the following good properties:
    \begin{itemize}\item With this definition it is true that $\Psi_\max$ is a state of maximum entropy;
    \item Entropy defined this way agrees with the Gibbons-Hawking definition
    $$S_\tot= \frac{A}{4G}+S_\vis$$
    {\it up to an additive constant independent of the state}, a point that we will return to shortly.  \end{itemize}  }
    
    So we have arrived at a microscopic definition of cosmological entropy with some good properties, by taking seriously the idea that what we want
    is to define the entropy of the universe {\it  from the perspective of an observer who lives in it}. 
    
    There are two flaws:
    \begin{itemize} \item As presented (and what I have said is equivalent to the construction in the paper \cite{CLPW}) this definition of entropy makes essential use of the 
    properties of de Sitter space, namely the existence of the symmetry generator $H$. So this definition does not carry over immediately to another spacetime.  
    \item The definition as given only agrees with Gibbons and Hawking up to an overall additive constant independent of the state. \end{itemize}  
    The reason that we are missing an additive constant in the entropy is that there was some renormalization in the definition.  In the quantum mechanical
    formula 
 \be\Tr\,\a=N \,\Tr\,\a\rho_\max\ee
 we threw away the factor $N$ (the Hilbert space dimension) because it is infinite in quantum field theory, and this leads to an additive renormalization of the entropy over which we have no control.\footnote{\label{onef}If we rescale the trace by $\Tr\to\lambda\Tr$ with $\lambda>0$, then to preserve the normalization condition
 $\Tr\,\rho=1$, we must rescale $\rho\to \lambda^{-1}\rho$.   Then $S(\rho)$ is shifted by $S(\rho)\to S(\rho)+\log\lambda$.}
 
 {The first problem can be cured by a simple change of perspective.   We consider an observer on a timelike geodesic in an arbitrary spacetime.
  Instead of the symmetries of the spacetime (since in general there are none), we consider the symmetries of the worldline itself.    An operator $\phi(\tau)$,
 where $\tau$ is the observer proper time, is not meaningful because there is no natural definition of $\tau$; it is only defined up to an overall additive constant.  
 So we again need to give the observer a clock and measure $\tau$ relative to this clock.   After a short discussion, this leads to an algebra $\A$ that generalizes
 what we said in de Sitter space to an arbitrary spacetime \cite{W}.}
 
 {To improve the definition for an observer in a general spacetime so that there is no undetermined additive constant,
  we need a good analog of the state $\Psi_\max$ in a general spacetime.     Here I have to go out on a limb.    There is an interesting candidate,
 but it is not well understood.    Hartle and Hawking \cite{HH} defined the ``no boundary'' state $\Psi_{\NB}$ as a generalization of the natural de Sitter invariant state $\Psi_\dS$.
   It is defined by a path integral on a spacetime that has ``no boundaries'' except the one where one wants to define the state.  I should stress that very little about this
 state is well established.    But a guess is that  the
 no boundary state, generalized to include the observer, has similar  properties to $\Psi_\max$: it  is a universal state of maximum entropy and the expectation of an operator in this state behaves as a trace.}
 
 If so,  we can give a universal definition of entropy for an observer
 interacting with horizons as 
 \be \label{hopfor} S(\rho)=-S(\rho|\sigma_{\rm{NB}}),\ee
 where $\sigma_{\rm{NB}}$ is the density matrix of the no boundary state, 
 $\rho$ is a completely general state of the observer algebra whose entropy we want to define,
 and $S(\rho|\sigma_{\rm{NB}})$ is the relative entropy between $\rho$ and  $\sigma_{\rm{NB}}$. 
 In any algebra with a trace, 
 relative entropy is defined by the same formula as in nonrelativistic quantum mechanics:
 \be S(\rho|\sigma)=\Tr\,(\rho \log\rho-\rho \log\sigma).\ee
This formula is invariant under rescaling of the trace,\footnote{The logic is as in footnote \ref{onef} though the conclusion is different.
$S(\rho|\sigma_\NB)$ is invariant under the rescaling $\Tr\to\lambda\,\Tr$, $\rho\to \lambda^{-1}\rho,$ $\sigma_\NB\to \lambda^{-1}\sigma_\NB$.} so it is not subject to the renormalization that we encountered in defining entropy directly.
  In the formula (\ref{hopfor}), I am proposing to use the Hartle-Hawking no boundary state similarly to the way  Wall used the Hartle-Hawking state of a black hole in studying the Second Law \cite{Wall}.
 
 Assuming we do {\it not} try to normalize the no boundary density matrix $\sigma_{\rm{NB}}$ but just take it directly from the gravitational   path integral,
 this is potentially a universal definition of cosmological entropy that agrees with the Gibbons-Hawking proposal with no missing constant.
   For some evidence for the proposal, see \cite{AKU}.
     A physical motivation  to not try to normalize
 the no boundary density matrix is that an observer does not know how to do so.    The observer can (hopefully!) study the no boundary density matrix in the spacetime  that
 the observer lives in, but the normalization of the no boundary density matrix would be dominated by the contribution of some other spacetime that the observer might not know about.   It is plausible that the no boundary state is actually unnormalizable and in that case we certainly should not try to normalize it.

In concluding, I will discuss some criticisms of this approach and my opinions about them.    First, some physicists object to what they regard as artificially
adding an observer to the theory.   My view is the following.   I assume we are studying a theory that is rich enough that an observer can be modeled in this theory.
  For example, the theory may include the Standard Model and then we ourselves can be modeled in that theory.   In such a theory, I analyze not all states but only those
states in which an observer is present.    The reason to make this restriction is that since cosmological entropy is only defined relative to the presence of an observer,
we should  be prepared to take into account the presence of an observer when we try to discuss cosmological entropy.

Second, the model of the observer was highly idealized.    We
 assumed that the observer can measure arbitrary operators along the observer's worldline and we did not worry about the resources required to do so.    A more realistic  -- and inevitably more complicated -- model would involve
 modeling the observer's capabilities in some fashion.    I believe that the idealized model is the right one if one wishes to understand the meaning of the Gibbons-Hawking
 proposal for cosmological entropy.    After all, one can think of entropy as the information that could be extracted from the state, but has not yet been extracted.   Any restriction
 on the observer's capabilities will reduce the information that the observer can extract from the state, so the entropy from such an observer's standpoint
 will be less than the Gibbons-Hawking value.   As a simple
 example of this, putting an upper bound on the energy available to the observer can be shown to reduce the entropy.

\vskip1cm
 \noindent {\it {Acknowledgements}}  
  Research supported in part by NSF Grant PHY-2514611.
 \bibliographystyle{unsrt}

\end{document}